\documentclass[pre,showpacs,preprint]{revtex4}
\usepackage{amssymb,graphicx,amsbsy}
\usepackage{bm}

\begin{document}

\title{Analytic results for the percolation transitions of the enhanced
binary tree}
\author{Petter Minnhagen}
\email[Corresponding author, E-mail: ]{Petter.Minnhagen@physics.umu.se}
\affiliation{Department of Physics, Ume{\aa} University, 901 87 Ume{\aa},
Sweden}
\author{Seung Ki Baek}
\affiliation{Department of Physics, Ume{\aa} University, 901 87 Ume{\aa},
Sweden}
\begin{abstract}
Percolation for a planar lattice has a single percolation threshold, whereas
percolation for a negatively curved lattice displays two separate thresholds.
The enhanced binary tree (EBT) can be viewed as a
prototype model displaying two separate percolation thresholds.
We present an analytic result for the EBT model which gives
two critical percolation threshold probabilities,
$p_{c1}=\frac{1}{2}\sqrt{13}-\frac{3}{2}$ and $p_{c2}=1/2$,
and yields a size-scaling exponent $\Phi =\ln
\left[\frac{p(1+p)}{1-p(1-p)}\right]/\ln 2$. It is inferred that the two
threshold
values give exact upper limits and that $p_{c1}$ is furthermore exact. In
addition, we argue that $p_{c2}$ is also exact. The physics of the model and
the results are described within the
midpoint-percolation concept: Monte Carlo simulations are presented for the
number of boundary points which are reached from the midpoint, and the
results are compared to the number of routes from the midpoint to the
boundary given by the analytic solution. These comparisons
provide a more precise physical picture of what happens at the transitions.
Finally, the results are compared to related works, in particular, the Cayley
tree and Monte Carlo results for hyperbolic lattices as well as earlier
results for the EBT model. It disproves a conjecture that the EBT has an
exact relation to the thresholds of its dual lattice.
\end{abstract}

%64.60.ah 	Percolation
%05.70.Jk 	Critical point phenomena

\maketitle

\section{Introduction}

\begin{figure}
\includegraphics[width=0.20\textwidth]{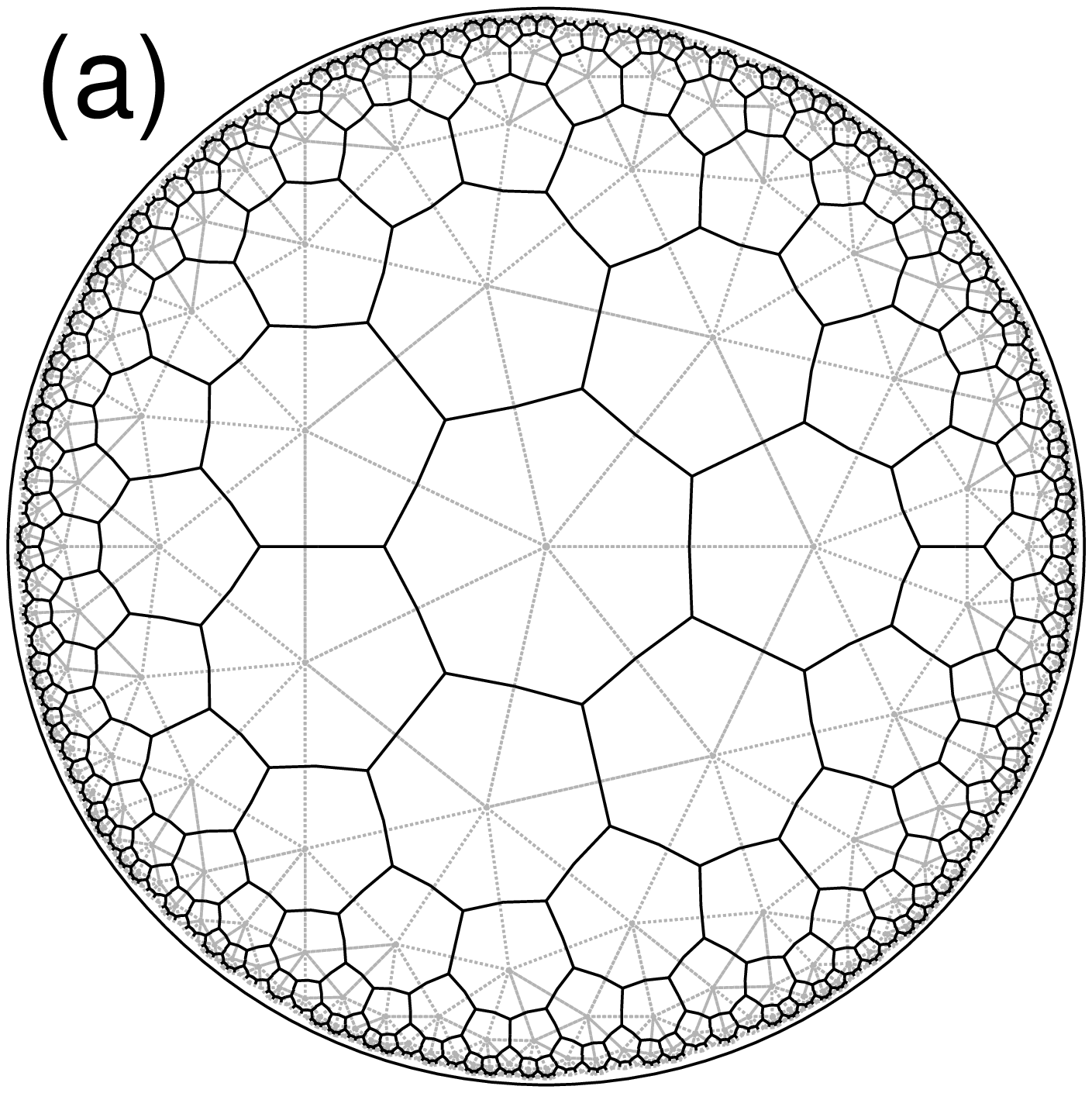}
\includegraphics[width=0.20\textwidth]{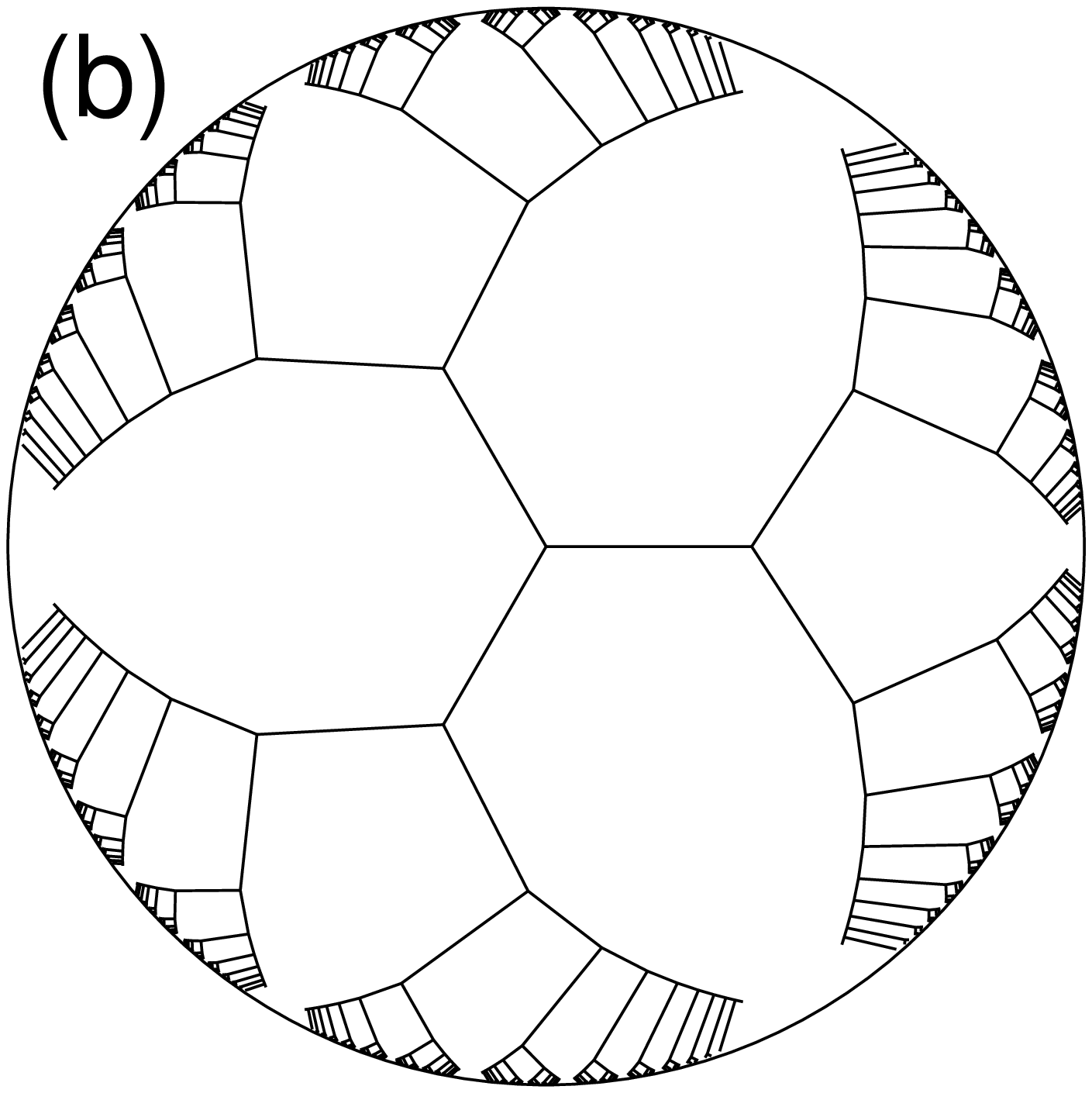}
\includegraphics[width=0.20\textwidth]{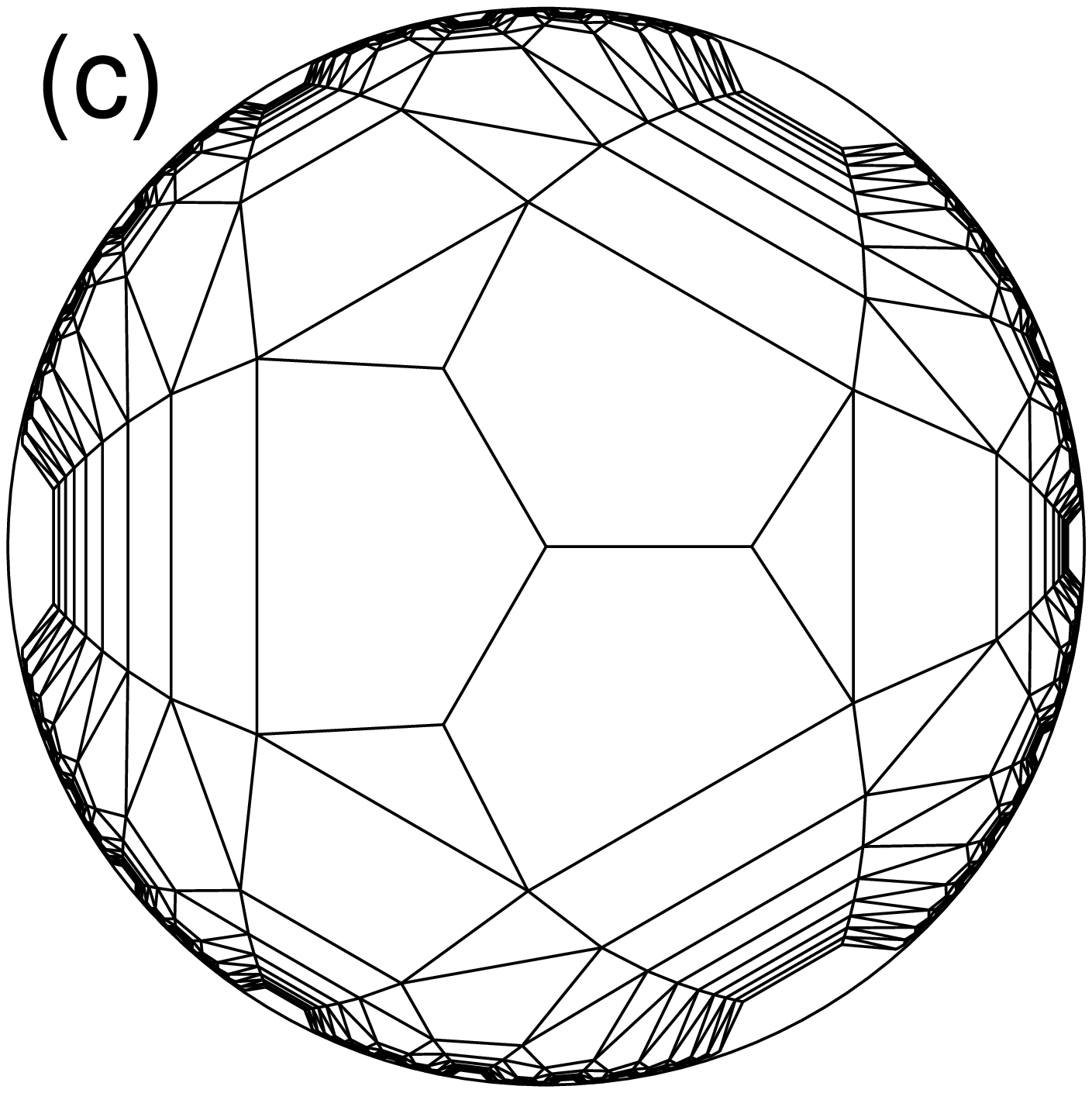}
\caption{Hyperbolic lattice structures drawn on the Poincar\'{e} disk. (a)
The solid lines show a heptagonal lattice, $\{7,3\}$, and the dotted lines
show its dual, $\{3,7\}$. (b) A tree structure, represented as
$\{\infty,3\}$, up to the maximum level $L=10$. (c) The enhanced tree obtained
from (b) by connecting the tree branches.}
\label{fig:poin}
\end{figure}

The percolation transition is one of the main examples of a critical
phenomena~\cite{stauffer,chris}, and there have been extensive studies on
lattices in Euclidean space, in particular, various types of planar lattices.
An interesting and somewhat less studied aspect
of the percolation phenomena is that, while a planar lattice possesses a
single percolation threshold, a hyperbolic lattice with a constant negative
curvature has two different thresholds~\cite{baek}.
Figure~\ref{fig:poin}(a) gives an
example of a hyperbolic lattice with Schl\"afli symbol $\{7,3\}$ as well as
its dual, $\{3,7\}$, where $\{m,n\}$ means that the number of regular
$m$-gons meeting at each vertex inside the structure is $n$.
One characteristic feature of negative
curved lattices is that the number of boundary points is a finite fraction
of the total number of lattice points. The fact that the number of surface
points is a finite fraction is intimately connected to the existence of two
thresholds: the probability of reaching \emph{one} of the boundary point
from the midpoint becomes finite at occupation probability $p=p_{c1}$
in the infinite-size limit,
whereas the probability to reach a finite \emph{fraction} of the boundary
points becomes finite at $p_{c2}$. This midpoint-percolation description
offers a conceptually simple way of understanding the physics at the two
thresholds. The hyperbolic lattices have, in comparison with the planar
lattices, two complications: one is that there are few known exact results
for the actual values of the thresholds and for the critical properties
associated with them. The other is that, since the number of lattice points
grows rapidly with the distance from the midpoint, it is hard to obtain very
precise results numerically by using standard size scaling in conjunction
with Monte Carlo techniques. The enhanced 
binary tree (EBT) model alleviates these difficulties to some extent.

The EBT was introduced in the present context by Nogawa and
Hasegawa~\cite{nogawa} as a
simplified model for a negatively curved lattice. Like the usual
Cayley tree, it retains the key feature that the boundary
points are a finite fraction of the total number of lattice points. However,
the EBT model has an advantage over the usual hyperbolic lattices because
its smaller fraction of boundary points, which makes it easier to obtain
numerical results for larger sizes. Even so, recent conflicting numerical
results for the EBT model reflect the difficulty inherent in obtaining
reliable numerical results for hyperbolic lattices~\cite{com,reply}. Hence it
would be useful to have exact estimates for a prototype model, which can be
used as a benchmark for the various numerical techniques (see
Refs.~\cite{sykes,kesten,wier,scu,ziff06,scu06,boettcher,berker}
for some existing exact threshold calculations).
It is then fortunate that, as shown here, the EBT model itself
possesses analytically accessible results.

\begin{figure}
\includegraphics[width=0.30\textwidth]{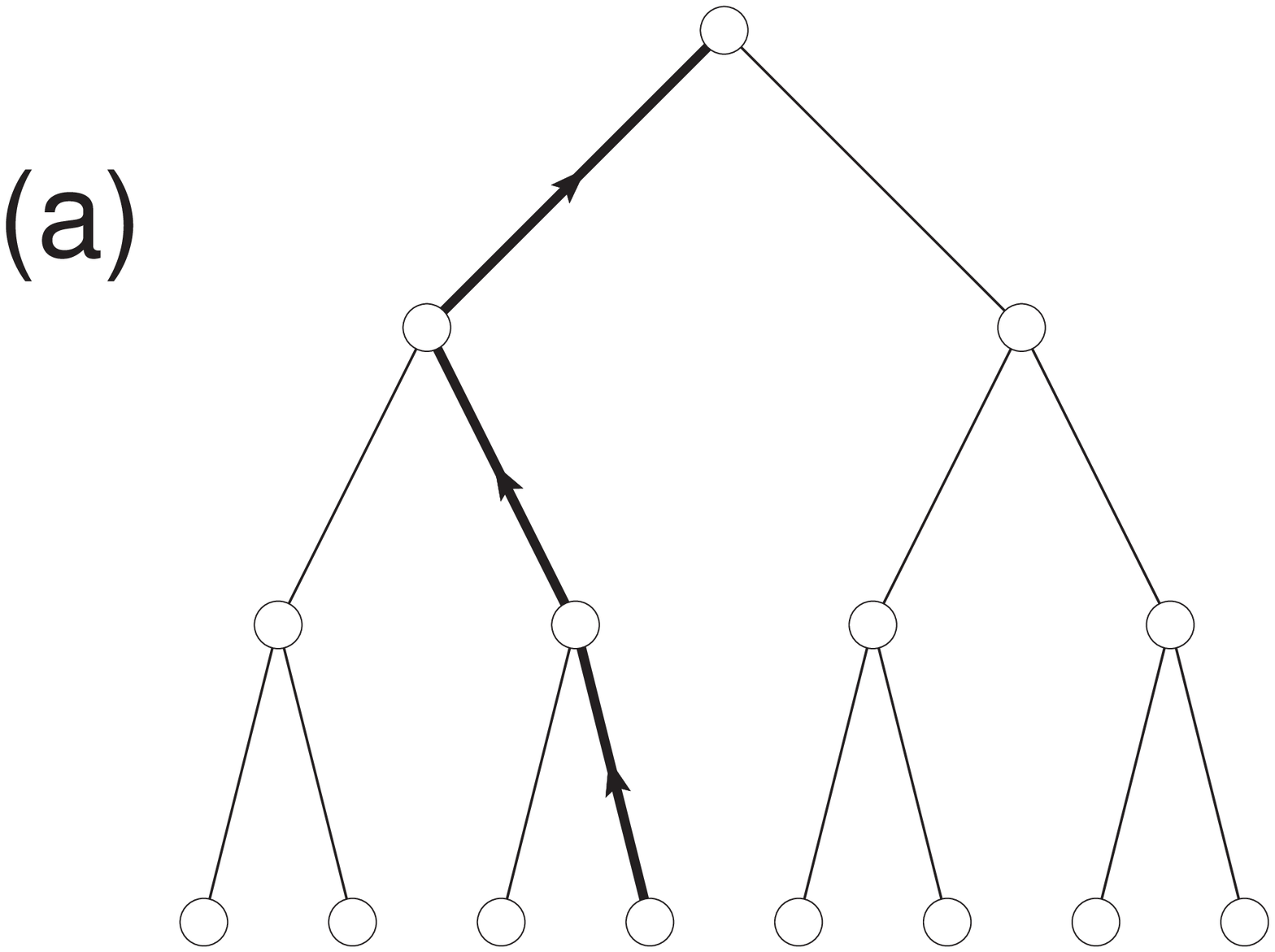}
\includegraphics[width=0.30\textwidth]{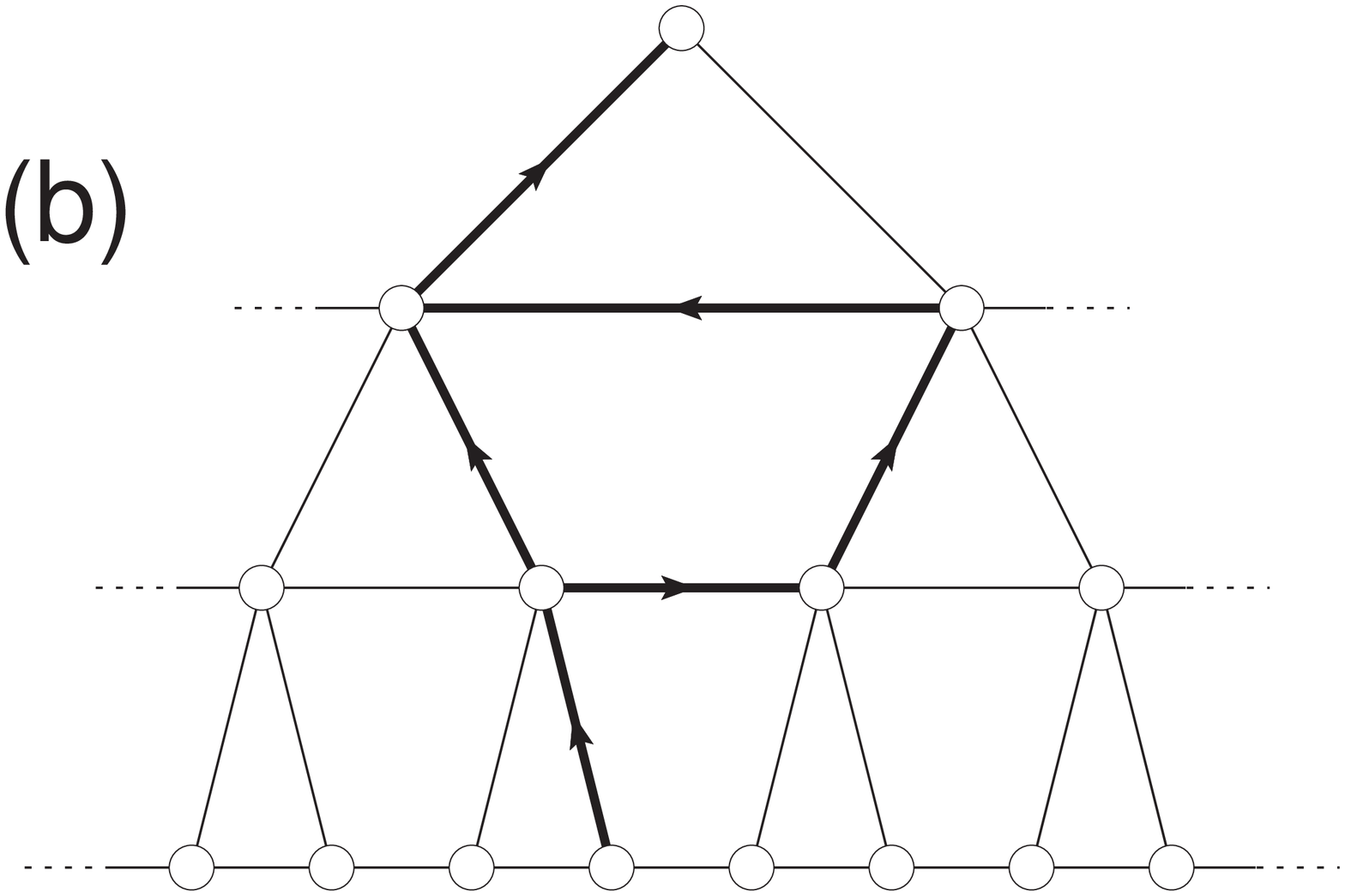}
\includegraphics[width=0.30\textwidth]{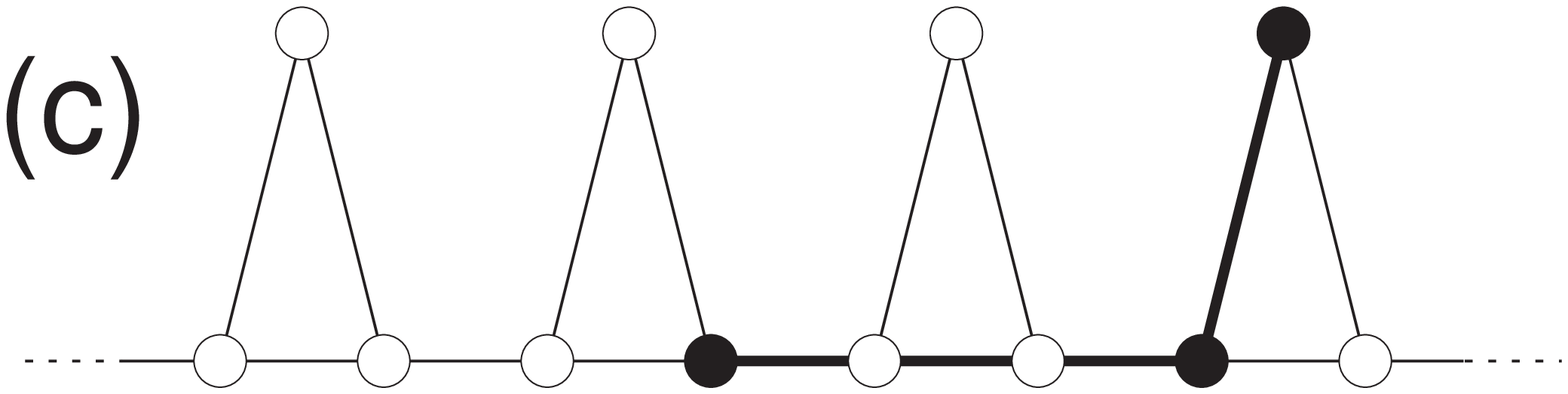}
\caption{Schematic representations of (a) the binary tree and (b) EBT.
Arrows show examples of routes from the surface to the midpoint. In panel
(a), there is only one possible route from a surface point to the top,
whereas in panel (b), there are several because the tree branches are
connected. (c) A horizontal layer with a rightward tagged route. The three
black dots denote the first point of the route on level $l$, the last point
of route on level $l$, and the first point of the route on level $l-1$,
respectively.}
\label{fig:latt}
\end{figure}

The EBT structure is obtained
from the usual binary tree by adding horizontal
cross links together with a periodic boundary in the horizontal
direction, as shown in Figs.~\ref{fig:poin}(c) and \ref{fig:latt}(b). As is
apparent by comparing the hyperbolic lattice in Fig.~\ref{fig:poin}(a) with
the EBT in Fig.~\ref{fig:poin}(c), they have similar structures: the
essential shared properties are first that the number of
lattice points on the surface, $B$, is a finite fraction of the total number
of lattice points, $\Omega$, in the limit $\Omega \rightarrow \infty$, and
second that there are horizontal paths leading all the way around the
lattices. Comparing the binary tree and the EBT in Figs.~\ref{fig:latt}(a)
and \ref{fig:latt}(b) shows that the crucial difference is that the distinct
branches for the binary tree are horizontally linked together in the EBT.

\section{Analytic Result}

The method used to find solutions for the EBT model can be illustrated
using the binary tree. We pick an arbitrary point on the surface and ask
how many routes there are which connect this particular point
to the midpoint. For the binary tree the solution is trivial: as
seen from Fig.~\ref{fig:latt}(a), there is only one possible route.
But in order to reach
the top, all the links along this route have to be in place. The chance for
a link to be in place is given by the occupation probability $0\leq p\leq
1$. If the number of steps needed to reach the top is $L$, then the
probability to get to the top is $P=p^{L},$ which obviously goes to zero for 
$L\rightarrow \infty$ unless $p=1$. So, if one picks an arbitrary surface
point, there is for $p=1$ a finite probability for the existence of a
connected route to the midpoint. This means that all the surface points have
a finite probability to be connected to the midpoint. Hence at $p_{c2}=1$, a
finite fraction of the surface point is connected to the midpoint (in this
trivial case, this fraction is 1). The number of surface points is $B=2^{L}$,
and one can alternatively ask what is the chance for finding a connection
from the midpoint to just one of the surface points. This is just $P\cdot
B=p^{L}2^{L}=(2p)^{L}$,
which becomes finite for $p=1/2$ and defines the first
percolation threshold $p_{c1}=1/2$. Thus the binary tree possesses two
separate thresholds: the lower one $p_{c1}$ defined as the first appearance
of a connection from the midpoint to the surface and the second one,
$p_{c2}$, defined as the first appearance of midpoint to surface connections
which reaches a finite fraction of the surface points.

We apply the same method to obtain the thresholds for the EBT case: two
possible routes from a surface point to the midpoint are drawn in
Fig.~\ref{fig:latt}(b). The difference with the binary tree is that a route
for the EBT can jump between the various tree branches. Thus there are now
many possible routes and the chance that one of them goes all the way to the
top for a given $p$ increases. As a consequence, the threshold $p_{c2}$
decreases to a value smaller than 1.

We first make a classification of possible routes to the midpoint from the
chosen starting point on the surface. A route classification is defined as
an ordered sequence of lattice points, route-markers, given by
$(l,d_{l},s_{l})$: if we start from a route-marker point at distance $l$
from the midpoint then the next route-marker point is identified by
$d_{l}=\pm 1$, which means that the next is left or right along the
horizontal level $l$, together with $s_{l}$ which denotes the distance in
lattice points to the next route-marker. At this route marker, the route
changes level so that the following route-marker is the closest lattice
point on level $l-1$, which means that this point is the first point on
the route on level $l-1$.
From this first point on $l-1$, we then identify the next two
route-markers by $(l-1,d_{l-1},s_{l-1})$, which gives the last point on the
route on level $l-1$, as well as the first on level $l-2$. The route
classification is illustrated in Fig.~\ref{fig:latt}(c). Next we restrict
the routes to all the routes which do not contain any backstep, where a
backstep is a step which increases the distance to the midpoint. Thus we
only consider steps which are either horizontal or upwards in
Fig.~\ref{fig:latt}(b). Also note that if $s_{l}=0$, the route goes directly
to the closest point on level $l-1$. Thus a complete sequence of $l$
route-marker labels $(l,d_{l},s_{l})$ gives a class of routes from the
surface point to the midpoint.

The next step is to consider how we get from the starting point on level $l$
to the end point on level $l-1$ for one specific link configuration. As
one moves horizontally on level $l$, one arrives at the possible
route-markers in consecutive order. Some of the lattice points are linked to
the level above, while others are not. Only the ones which are linked are
possible route-markers. These possible route-marker points can be enumerated
in consecutive order $s_{l}^{(k)}$ with $k=1,2,...$, so that
$(l,d_{l},s_{l}^{(k)})$ enumerates all the possible route classes we can
have when starting from the first point on level $l$ going horizontally in
the direction $d_{l}$ to the next route marker, which connects to the level
$l-1$. Out of these possibly many classes one picks one specific, the one
with the shortest horizontal distance $(l,d,s_{l}^{(1)})$. By choosing the
shortest route-marker distance on every level one obtains a class of tagged
routes from the point on the surface to the midpoint. The probability for
one such tagged route, denoted by $j$, is $P_{j}^{\rm (tag)}=\prod_{l=0}^{L}
P^{\rm (tag)}(l,d_{l},s_{l}^{(1)})$ and the average number of such tagged
routes, which reach the midpoint, is $r^{\rm (tag)}=\sum_{j}P_{j}^{\rm (tag)}$.
Here $j=1...2^L$ enumerates the possible tagged routes which correspond to
the possible sequences $d(1),d(2),...d(L)$. 
The point is that $r^{\rm (tag)}$ can be explicitly calculated: starting
from the first route-marker on level $l$ in Fig.~\ref{fig:latt}(c), the
chance for getting to the closest point on the level $l-1$ above, when
moving to the left, is $b(p)=p+(1-p)p^{2},$ because there are two routes
leading to the closest point. In the same way, the chance to end at the
second closest is $a(p)b(p)$ where $a(p)=p^{2}(1-p)^{2}$, the third
$a^{2}(p)b(p)$ and to point number $i$ is $a^{i-1}(p)b(p)$. This also means
that the chance of ending at one of the points on the level above, when
moving in the left direction starting from a right triangle corner is
\begin{equation}
Q_{\rm left}(l)=\sum_{\nu =1}^{N(l-1)}a^{\nu -1}(p)b(p)=b(p)\frac{
1-a^{N(l-1)}(p)}{1-a(p)},
\end{equation}
where $N(l-1)=2^{l-1}$ is the number of points on the $l-1$ level.
In the rightward direction, like the route drawn in Fig.~\ref{fig:latt}(c),
we first have to move one step to the right with probability $p$ and the
result becomes
\begin{equation}
Q_{\rm right}(l)=p~b(p)\frac{1-a^{N(l-1)}(p)}{1-a(p)}.
\end{equation}
The probability for a tagged route with no backsteps is then given by
$\prod_{l=1}^{L}Q_{d(l)}(l)$ where $d(l)=\pm 1$ defines a route to
the left or right on that particular $l$-level. The average number of
such tagged routes is hence 
\begin{eqnarray}
r^{\rm tag} &=& \prod_{l=1}^{L} \left[ Q_{\rm left}(l)+Q_{\rm right}(l)
\right] \nonumber\\
&=& \left[\frac{(1+p)b(p)}{1-a(p)} \right]^{L}\prod_{l=1}^{L} \left[
1-a^{N(l-1)}(p) \right] =
\left[ \frac{p(1+p)}{1-p(1-p)} \right]^{L}\prod_{l=1}^{L} \left[
1-p^{2}(1-p)^{2^{l}} \right],
\label{rtag}
\end{eqnarray}
which means that 
\begin{equation}
c \left[ \frac{p(1+p)}{1-p(1-p)} \right]^{L}\leq r^{\rm tag}\leq
\left[ \frac{p(1+p)}{1-p(1-p)} \right]^{L},
\label{size}
\end{equation}
where $0<c\leq 1$ is a constant independent of $L$. Thus for
$\frac{p(1+p)}{1-p(1-p)}<1$, the average number of tagged routes without
backsteps vanishes in the limit $L\rightarrow \infty $. This happens for
$p<1/2$ which gives the estimate $p_{c2}=1/2$: at this threshold the
probability for having a tagged route from an arbitrary surface point to the
midpoint vanishes. In contrast, for $p>1/2$ there is a nonzero probability
that a finite fraction of the surface points are connected to the midpoint.
Since the number of surface points is $B=2^{L}$, this also means that, when
$Br^{\rm tag} $ becomes finite, there is a finite probability that that one of
the surface points is connected to the midpoint. This condition is given by
$\frac{p(1+p)}{1-p(1-p)}=1/2$ and has the solution
$p_{c1}=\frac{1}{2}\sqrt{13}-\frac{3}{2}\approx 0.30278$. Since the
calculation is based on a particular subclass of possible routes, these
results are either upper bounds or exact.

\section{Exactness Argument}

\begin{figure}
\includegraphics[width=0.60\textwidth]{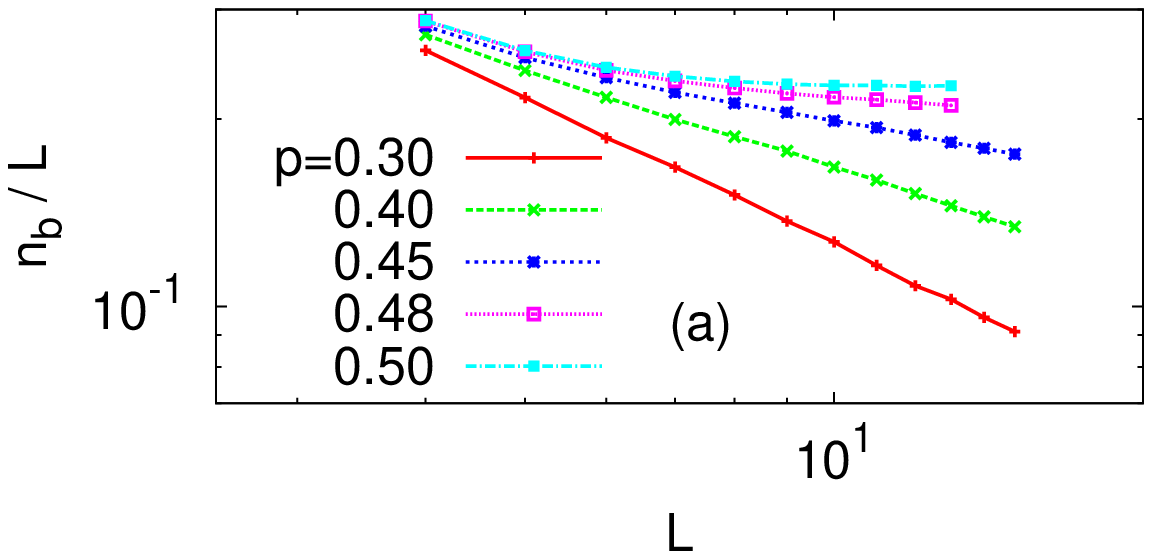}
\includegraphics[width=0.60\textwidth]{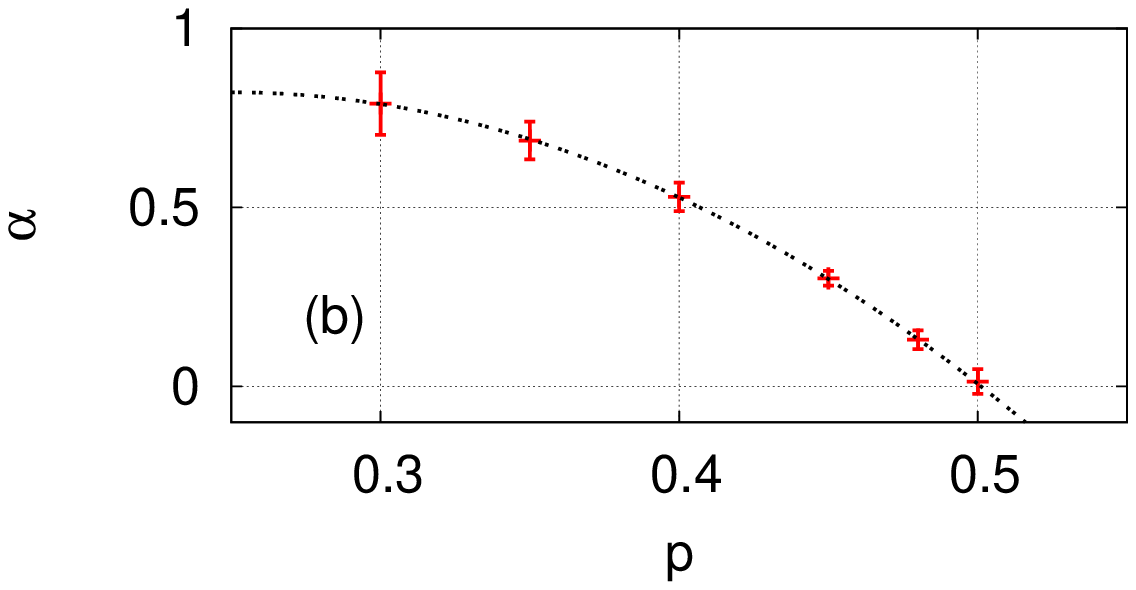}
\caption{(Color online) The minimum number of backsteps $n_b$ for a
nonzero-backstep path which connects from bottom to the top.
(a) The data is
plotted as $n_b/L$ against $L$ in a log-log plot for a sequence of fixed
$p$ values. For smaller $p$, the data shows that $n_b/L$ vanishes as a power
law $n_b/L\propto L^{-\alpha}$ with increasing $L$. However, for $p\approx
0.5$, it instead goes to a constant. This change of behavior signals
a phase transition. (b) The numerically determined power-law index $\alpha$
plotted as a function of $p$. The dotted line presents a second-order
polynomial fit for eyes guidance. Extrapolation for the data points
suggests that $\alpha$ vanishes at $p=0.50(1)$.
}  
\label{fig:frac}
\end{figure}

The tagged route thresholds values for $p_{c1}$ and $p_{c2}$ are by
construction upper bounds. The question is then if they also are the exact
values. A heuristic argument for the exactness of $p_{c2}$ goes as follows:
at the real $p_{c2}$ a unique percolating cluster is formed. The formation
of such a cluster means that the probability of finding paths, which
takes you to the top, suddenly increases. Such a
probability increase is, by the tagged-route argument, predicted to happen
at $p=0.5$. This suggests that the tagged-route prediction
$p_{c2}\leq 0.5$ can be sharpened to $p_{c2}=0.5$. 

In order to further probe the exactness, we numerically
calculate the minimum number of backsteps, $n_b$, needed for a
nonzero-backstep path to go from the bottom to the top.
Figure~\ref{fig:frac}(a) shows that $n_b/L$ goes to zero in the limit
$L\rightarrow\infty$ in the region below $p=0.5$ and furthermore that this
size scaling is a power law $n_b/L\propto L^{-\alpha}$.
Figure~\ref{fig:frac}(b) shows that the power-law index decreases with
increasing $p$ and
vanishes at a critical value $p_{c2}\approx 0.5$. Since this vanishing slope
is clearly in itself a critical behavior, it is plausible that the slope
vanishes at the exact threshold $p_{c2}\leq 0.5$. Again the formation of a
single percolating cluster at $p_{c2}$ is likely to be associated with a
sudden change of the properties of the routes which take you from the bottom
to the top. Since a sudden change in the number of backsteps is just such a
property, this suggests that $\alpha$ vanishes precisely at $p_{c2}=0.5$.
The numerical results are consistent with vanishing at $p=0.5$, but our
numerical precision is not high enough to rule out that it could also vanish
somewhere in the region $0.49<p<0.5$.

The fact that the minimum number of backsteps per level vanishes at least
below $p=0.49$ makes it possible to deduce that the tagged-route value for
the lower threshold $p_{c1}=\frac{1}{2}\sqrt{13}-\frac{3}{2}\approx 0.30278$
is exact. The reason is that a typical path from the bottom to the top will
always contain at least one section where the distance between two backsteps
along the path is proportional to $L$. This means that a typical path will
always contain sections without backsteps which connects levels separated
with a distance which goes to infinity in the limit $L\rightarrow\infty$.
The tagged-route argument predicts that such a path exists (does not exist)
with a finite probability for $p$ larger (smaller) than
$p_{c1}=\frac{1}{2}\sqrt{13}-\frac{3}{2}$.
The same argument tells you that $p_{c2}$ cannot be lower than the
$p$ value where $\alpha$ vanishes, which according to our numerical
extrapolation in Fig.~\ref{fig:frac}(b) gives $0.49< p_{c2} \leq 0.5$. Note
that the upper limit in this estimate is $0.5$ because the tagged-route
argument by itself gives an upper bound.

To sum up: we conclude that $p_{c1}=\frac{1}{2}\sqrt{13}-\frac{3}{2}$ is the
exact value and that $0.49< p_{c2}\leq 0.5$. Furthermore, we conjecture that
$p_{c2}=0.5$ is likewise exact based on plausibility arguments.
 
\section{Comparisons}

\begin{figure}
\includegraphics[width=0.45\textwidth]{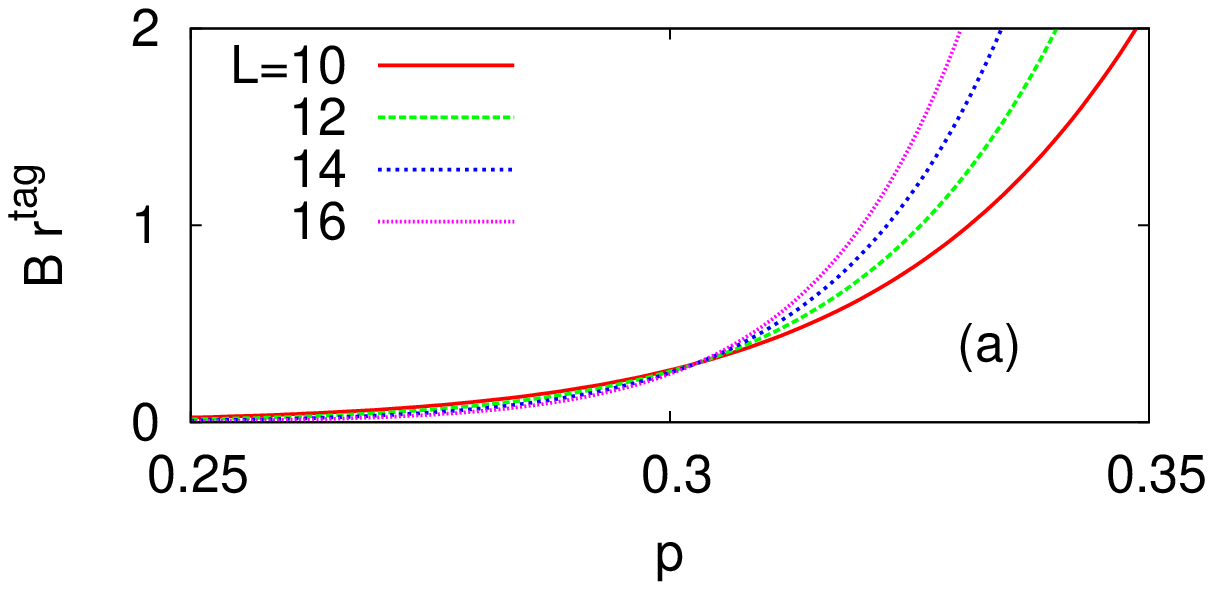}
\includegraphics[width=0.45\textwidth]{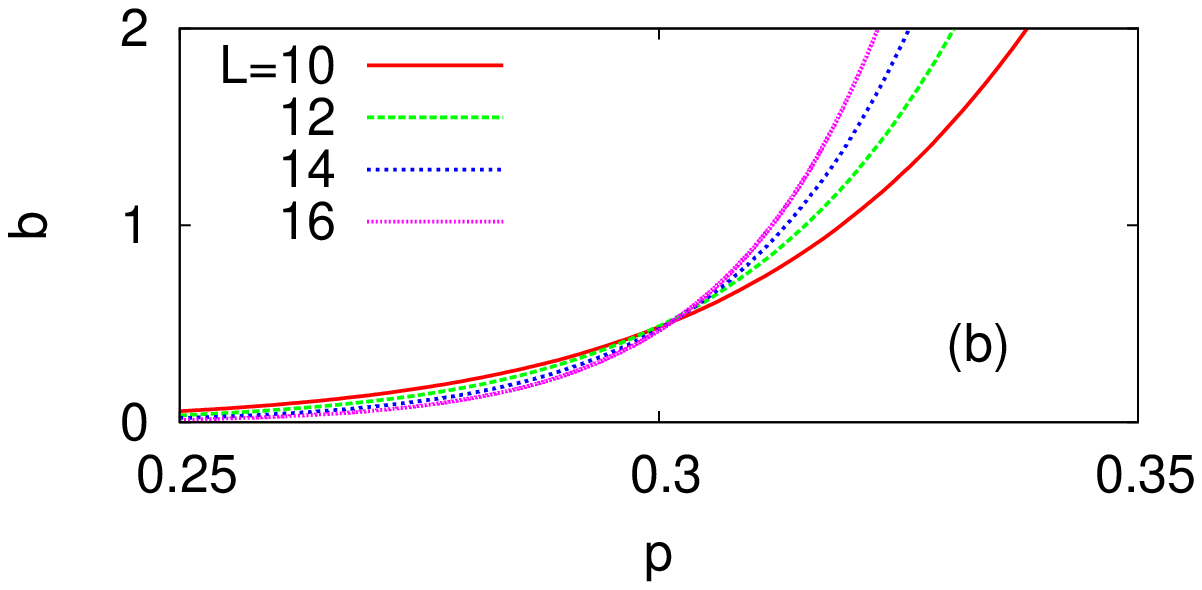}
\includegraphics[width=0.45\textwidth]{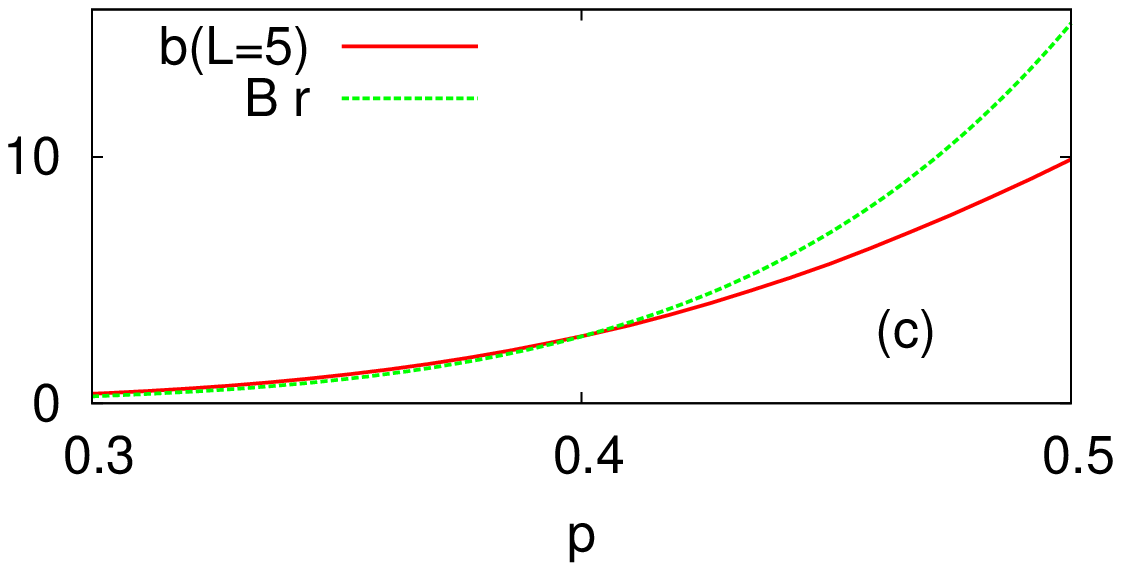}
\includegraphics[width=0.45\textwidth]{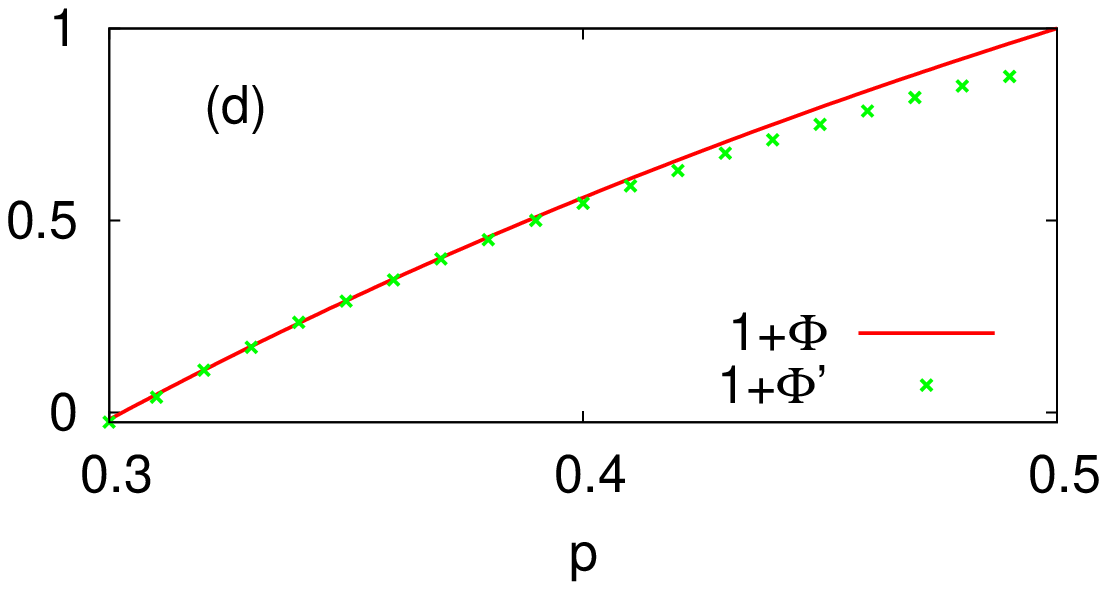}
\includegraphics[width=0.45\textwidth]{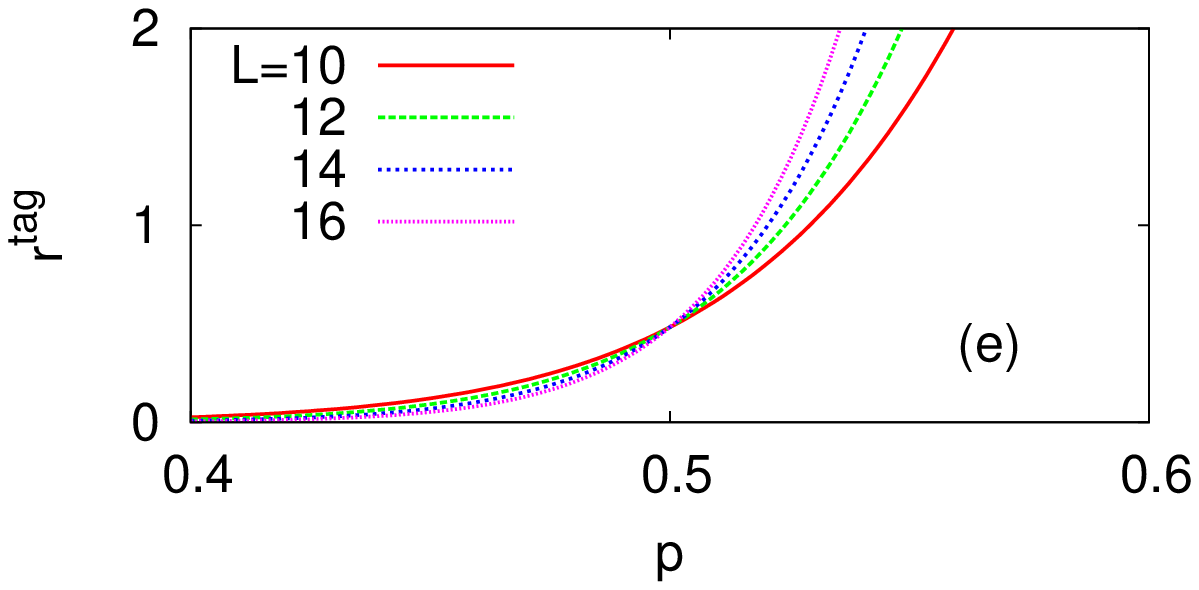}
\includegraphics[width=0.45\textwidth]{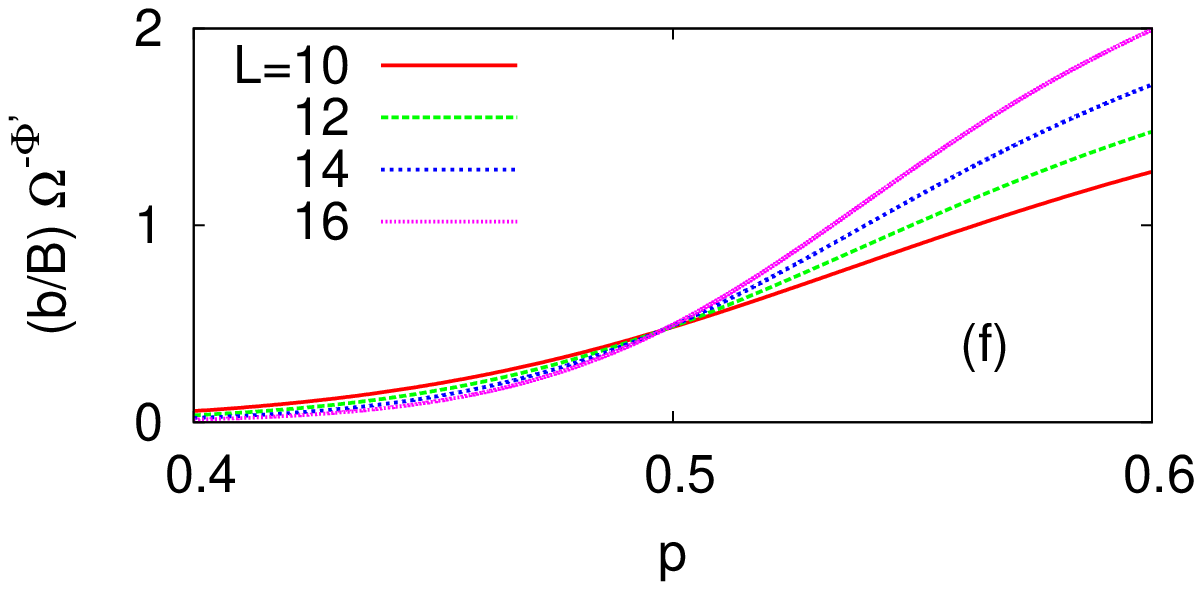}

\caption{(Color online) Comparisons between $r^{\rm tag}$, the average number of tagged routes
from an arbitrary surface point to the midpoint, and, $b$, the number of
surface points connected to the midpoint. $r^{\rm tag}$ is given by the
analytical solution and $b$ is obtained from numerical simulations. (a) The
number of tagged routes from the surface to the midpoint, $Br^{\rm tag}$ for
various sizes $L$. All curves cross in a single point at $p_{c1}\approx
0.303$. (b) The same plot for $b$. Note the strong resemblance and the
single crossing point at $p\approx 0.303(1)$. The panel (c) illustrates the
close connection between $Br^{\rm tag}$ and $b$. For $p\leq0.4$ the number
of routes and the number of connections are almost the same (the number of
routes is slightly smaller as explained in the text). However for $p\geq
0.4$, the number of routes exceeds the number of connected surface points.
The panel (d) illustrates this connection further
by comparing the size-scaling exponent $1+\Phi$ obtained analytically for
$r^{\rm tag}$ with the size scaling for $b$ obtained from (b). The
agreement is striking for $p\leq 0.4$. (e) $r^{\rm tag}$ for various sizes
with the crossing point at $p_{c2}=1/2$. (f) The corresponding crossing
point for $b/B$ obtained numerically. The crossing point is close to 0.50,
but the size-scaling exponent $\Phi$ is different. For $r^{\rm tag}$ the
value is $\Phi(p_{c2})=0$ but for $b$ it is $\Phi'(p_{c2})\approx -0.11$,
as explained in the text.}
\label{fig:comp}
\end{figure}
 
Figure~\ref{fig:comp} gives a comparison between the \emph{number} of tagged
routes connecting the midpoint with the surface, $Br^{\rm tag}$ given by
Eq.~(\ref{rtag}), and $b$, the number of surface points connected to the
midpoint: Fig.~\ref{fig:comp}(a) shows the number of tagged routes as a
function of $p$ for various sizes $L$. The curves for the different sizes
cross at the critical value
$p_{c1}=\frac{1}{2}\sqrt{13}-\frac{3}{2}\approx0.3028$, whereas
Fig.~\ref{fig:comp}(b) shows
the numbers of surface points which are connected to the midpoint, $b$.
These values are obtained from numerical simulations as described in
Refs.~\cite{baek} and \cite{com}. These curves also cross at $p_{c1}$ and the independent
numerical estimate in Ref.~\cite{nogawa}, $p_{c1}=0.304(1)$, is also in
excellent agreement with this result. Note that $r^{\rm tag}$ and $b$ are
two distinct quantities although closely related. They would be identical if
the connection between a surface point and the midpoint did always involve
just a single tagged route. As seen in Fig.~\ref{fig:comp}(c), this is true
to good approximation in the region $0.3<p<0.4$. For small $p$, there is a
tiny deviation because a single tagged route has a small but finite chance
of connecting more than one surface point to the midpoint. Thus if a surface
point and a midpoint is always connected by precisely one route, then the
total number of routes is slightly smaller than the number of connected
surface points. On the other hand, for larger $p$, many routes can be
connected to the same surface point, so that the number of routes instead
exceeds the number of connected surface points. This effects becomes
noticeable for $p>0.4$.
Figure~\ref{fig:comp}(c) demonstrates the strong connection between the
tagged routes and the connections between the midpoint and the surface. This
correspondence is further enforced by comparing the size
dependence of $r^{\rm tag}$ and $b/B$. For $r^{\rm tag}$ the size dependence
follows from Eq.~(\ref{size}) from which one obtains $r^{\rm tag}\sim \Omega
^{\Phi}$, where $\Omega \sim 2^{L}$ is the size of the system and $\Phi =\ln
\left[\frac{p(1+p)}{1-p(1-p)})\right]/\ln 2$. The exponent $\Phi$ is plotted in
Fig.~\ref{fig:comp}(b) and is compared to the corresponding exponent for
$b/B\sim \Omega^{\Phi ^{\prime}}$ obtained numerically from
simulations~\cite{com,baek}. As seen $\Phi\approx\Phi^{\prime}$ for $p<0.4$,
which again shows that in this interval the number of connected surface
points is closely the same as the number of tagged routes. However, as
noticed above, for larger values of $p$, the number of routes exceeds the
number of connected surface points. This has interesting consequences at the
second threshold $p_{c2}$: in Fig.~\ref{fig:comp}(e), the number of tagged
routes per surface point is plotted and the curves for the various sizes
crosses precisely at $p_{c2}=0.5$. This means that, at this threshold, the
size-scaling exponent for the tagged routes is $\Phi =0$ whereas the
numerically determined size-scaling exponent for the fraction of attached
surface points, $b/B$ is $\Phi ^{\prime }\approx -0.11$, as shown in
Fig.~\ref{fig:comp}(e). This means that, at $p_{c2}$, the number of tagged
routes which are attached to the same surface point decreases with size as a
power law with the exponent $\Phi ^{\prime }\approx -0.11$. 
One also notes that the threshold value determined numerically from $b/B$
is consistent with the exact value $p_{c2}=1/2$, but
with a larger uncertainty than the numerical determination of $p_{c1}$.

One question is whether or not the EBT and its dual lattice are related in
such a way that the lower threshold  $p_{c1}^{\rm (dual)}$ for the dual
lattice and the upper threshold for the EBT $p_{c2}$, sum to  $p_{c1}^{\rm
(dual)}+p_{c2}=1$, as was conjectured by Nogawa and Hasegawa~\cite{nogawa}.
Just as for the EBT, the lower threshold for the dual lattice can be
numerically determined to good precision and is given by
$p_{c1}^{\rm (dual)}=0.436(1)$~\cite{nogawa}. The determination of $p_{c2}\leq 1/2$
in this work shows that in fact $p_{c1}^{\rm (dual)}+p_{c2}\leq 0.936<1$,
and that consequently the conjectured exact relation is not valid.
This property is shared with the hyperbolic lattices. For example, $p_{c2}$
is for the $\{7,3\}$ lattice in Fig.~\ref{fig:poin}(a) numerically
determined as $p_{c2}\approx 0.72$ whereas the lower threshold for the dual
lattice $\{3,7\}$ is approximately $p_{c1}^{\rm (dual)}\approx
0.20$~\cite{baek}. This gives $p_{c1}^{\rm (dual)}+p_{c2}\approx 0.92$ and
illustrates the strong resemblance between the EBT model and the hyperbolic
lattices.

\section{Summary}
The values of the two percolation thresholds for the EBT model were
inferred using a two-step method: the first step involved an exact
calculation of a particular class of routes from the surface to the
midpoint. The second step involved a numerical calculation of the minimum
number of backsteps for nonzero-backstep routes connecting a surface point
to the midpoint:
when this number increases slower than $L$, the tagged-route probability
determines the percolation. This made it possible to infer the exact value
of $p_{c1}$. Arguments were presented which, in our opinion, give strong
evidence that $p_{c2}$ is also given exactly by the tagged-route probability.
One might suspect that there are lattices for which
complete analytical solutions are difficult to find, where the tagged-route
probability determines the percolation threshold. For example, the exact
results for the triangular, quadratic, and honeycomb lattices are readily
reproduced.

Hyperbolic lattices in general have two separate percolation thresholds and
both of them can be identified within the midpoint-percolation concept: at
the lower threshold, a finite number of the surface points are reached from
the midpoint and at the higher, a finite \emph{fraction}. The EBT can be
viewed as a simplified version of a hyperbolic lattice and it is, to our
knowledge, the first lattice of this type with two nontrivial thresholds
where at least one has been exactly determined. There are also two trivial
cases, i.e., the Cayley tree with $p_{c1}=1/2$ and $p_{c2}=1$ and the dual to
the Cayley tree which has $p_{c1}=0$ and $p_{c2}=1/2$. In addition, there are
related models which also possess two percolation
thresholds~\cite{boettcher,berker,hasegawa}. In Ref.~\cite{berker}, the two
thresholds for a scale-free hierarchical network were exactly obtained and
found to be $p_{c1}=0$ and $p_{c2}=5/32$, which seems superficially somewhat
reminiscent of the dual to the Cayley tree in the sense that there is no
nonpercolating phase for any nonzero $p$. Likewise, two percolation
thresholds were found for a certain Hanoi network in Ref.~\cite{boettcher} and
determined to high precision. In this case, both thresholds are nonzero just
as for the EBT. However, the relation between the scale-free hierarchical
network and the Hanoi network, on the one hand, and the percolation
transitions for a hyperbolic lattice, on the other, is an interesting
question which remains to be elucidated.

Apart from being a good proxy of a hyperbolic lattice, the EBT has
interesting intrinsic properties as illustrated by the
size scaling and the close relation between the tagged routes and the
surface-to-midpoint connections. Consequently, it might turn out to be a
useful model for future studies of percolation on hyperbolic lattices, as
well as for percolation in general.

\acknowledgments
We acknowledge the support from the Swedish Research Council
with the Grant No. 621-2002-4135.
This research was conducted using the resources of High Performance
Computing Center North (HPC2N).

%\bibliographystyle{revtex}
%\bibliography{ebp}

\end{document}